\documentclass[12pt,a4paper]{article}
\usepackage{amsmath}
\usepackage{graphics}

\newcommand{\hepth}[1]{{\tt hep-th/#1}}
\newcommand{\plb}[3]{Phys.Lett. {\bf B#1} (#2) #3}

\newcommand{\ep}{\varepsilon} 

\newcommand{\nn}{\nonumber}

\newcommand{\p}{\vspace{6pt}\noindent}
\advance\textheight by \topskip
\makeatletter

\@addtoreset{equation}{section}
\def\section{\@startsection {section}{1}{\z@}{-8.5ex plus -1ex minus
 -.2ex}{3.3ex plus .2ex}{\large\bf}}
\def\subsection{\@startsection{subsection}{2}{\z@}{-3.25ex plus
 -1ex minus -.2ex}{1.5ex plus .2ex}{\bf}}
\def\subsubsection{\@startsection{subsubsection}{3}{\z@}{-3.25ex plus%
 -1ex minus -.2ex}{1.5ex plus .2ex}{\sl}}

\begin{document}
\begin{titlepage}
\vspace*{-2cm}
\begin{flushright}

 hep-th/0305022 \\

\end{flushright}

\vspace{0.3cm}

\begin{center}
{\Large {\bf Classically integrable field theories with defects }}\\
\vspace{1cm} {\large P.\ Bowcock\footnote
{\noindent E-mail: {\tt peter.bowcock@durham.ac.uk}}, E.\
Corrigan\footnote{\noindent E-mail: {\tt ec9@york.ac.uk}} and C.\ Zambon
\footnote{\noindent E-mail: {\tt cz106@york.ac.uk}} }\\
\vspace{0.3cm} {${}^a$}\ \em Department of Mathematical
Sciences\\ University of Durham\\ Durham DH1 3LE, U.K.\\
\vspace{0.3cm} {${}^{b\ c}\ $\em\it Department of
Mathematics \\ University of York\\
York YO10 5DD, U.K. }\\ \vspace{1cm}
{\bf{ABSTRACT}}
\end{center}

\begin{quote}
Some ideas and remarks are presented concerning a possible
Lagrangian approach to the study of internal boundary conditions
relating integrable fields at the junction of two domains. The
main example given in the article concerns single real scalar
fields in each domain and it is found that these may be free, of
Liouville type, or of sinh-Gordon type.
\end{quote}

\vfill
\end{titlepage}

\section{Introduction}
\label{s:intro}

Recently, there has been some interest in the study of integrable
classical or quantum field theories restricted to a half-line, or
interval, by imposing integrable boundary conditions, see for example
\cite{Fring93,Ghosh94a, Ghosh94b, Cor94a, Cor95, MacI95, Bow95}.
The simplest
situation, which is also the best
understood, contains a real self-interacting scalar field $\phi$ with
either a periodic ($\cos$), or non-periodic ($\cosh$) potential.
The sinh-Gordon model can be restricted to the left
half-line $-\infty\leq x\leq 0$, without losing integrability, by
imposing the boundary condition
\begin{equation}\label{bc}
 \left. \partial_x\phi\right|_{x=0}=\frac{\sqrt{2}m}{\beta}
 \left(\ep_0e^{-\frac{\beta}{\sqrt{2}}\phi(0,t)}-
\ep_1e^{\frac{\beta}{\sqrt{2}}\phi(0,t)}\right) ,
\end{equation}
where $m$ and $\beta$ are the bulk mass scale and coupling constant,
respectively, and $\ep_0$ and $\ep_1$ are two additional parameters
\cite{Ghosh94a, MacI95}.  This set of boundary conditions
generally breaks the reflection symmetry $\phi\rightarrow -\phi$
of the  model although the symmetry is explicitly preserved when
$\ep_0=\ep_1\equiv\ep$. The restriction of the sinh-Gordon model
to a half-line is a considerable complication, and renders the
model more interesting than it appears to be in the bulk. This is
because there will in general be additional states in the spectrum
associated with the boundary, together with a set of reflection
factors compatible with the bulk S-matrix
(see \cite{ Ghosh94a, Ghosh94b, Cor95, Zam79}). The
weak-strong coupling duality enjoyed by the bulk theory emerges in
a new light \cite{Cor97, Chen00, Cor00}.

\p In this article a slightly different situation is explored.
There is no reason in principle why the point $x=0$ should not be
an internal boundary linking a field theory in the region $x<0$
with a (possibly different) field theory in the region $x>0$. The
quantum version of this set up has been examined before and
imposing the requirements of integrability was found to be highly
restrictive. This sort of investigation was pioneered by Delfino,
Mussardo and Simonetti some years ago \cite{Delf94}, and there has
also been some recent interest \cite{Castro-Alvaredo2002,
Mintchev02}. However, the objective of this article is to explore
a Lagrangian version of this question and derive the conditions
linking the two field theories at their common boundary. This
situation does not appear to have been discussed previously,
although the results turn out to be interesting and reminiscent of
some earlier work by Tarasov \cite{Tarasov91}.

\p
Internal boundary conditions will be
referred to as
`defect' conditions.

\p
Integrability in the
bulk sinh-Gordon model requires the existence of conserved quantities
labelled by odd spins $s=\pm 1, \pm 3, \dots$, and some of these should
survive even in the presence of boundary conditions. Since boundary
conditions typically violate translation invariance, it is expected that
the `momentum-like' combinations of conserved quantities will not be
preserved. However, the `energy-like' combinations, or some subset of them,
might remain conserved, at least when suitably modified (see \cite{Ghosh94a}
for the paradigm). As was the case for the theory restricted to a half-line,
the spin three charge already supplies the most general restrictions on
the boundary condition. The Lax pair approach developed in
\cite{Bow95} can be adapted to this new context and used to re-derive the
boundary conditions, thereby demonstrating that the preservation of
higher spin energy-like charges imposes no further restrictions on the
boundary conditions.
This will be discussed below in section 4.

\p
The starting point for the discussion is the Lagrangian density for the
pair of real scalar fields $\phi_1,\, \phi_2$:
\begin{eqnarray}\label{Lagrangian}
{\cal L}&=&\theta (-x)\left(\frac{1}{2}(\partial\phi_1)^2 -V_1(\phi_1)\right)
+\theta(x)\left(\frac{1}{2}(\partial\phi_2)^2 -V_2(\phi_2)\right)\\
&&\ \ \ \ \ \ \ \  +\delta(x)\left(\frac{1}{2}(\phi_1\partial_t\phi_2-
\phi_2\partial_t\phi_1)-{\cal B}(\phi_1,\phi_2)\right),\nonumber
\end{eqnarray}
in which the bulk potentials $V_1,\, V_2$ depend only the fields
$\phi_1,\, \phi_2$, respectively, while the boundary potential
${\cal B}$ depends on the values of both fields at the boundary $x=0$.
The part of the boundary term depending on the time
derivatives of the fields is not the most general possibility. However,
although excluding terms of higher order in time derivatives, it is
sufficiently general for present purposes.
The field equations and associated boundary conditions are
\begin{eqnarray}\label{fieldequations}
\partial^2\phi_1 &=& -\frac{\partial V_1}{\partial\phi_1},\ \ \ x<0\\
\partial^2\phi_2 &=& -\frac{\partial V_2}{\partial\phi_2},\ \ \ x>0\\
\partial_x\phi_1 - \partial_t\phi_2 &=&-
\frac{\partial{\cal B}}{\partial\phi_1},\ \ \ x=0\\
\partial_x\phi_2 - \partial_t\phi_1 &=&
\phantom{-}\frac{\partial{\cal B}}{\partial\phi_2},\ \ \ x=0.
\end{eqnarray}

\section{Consequences of the spin three conservation law}

For a single real scalar field $\phi$ in the bulk,
the spin three densities satisfy
\begin{equation}\label{spinthree}
\partial_\mp T_{\pm 4}=\partial_\pm \Theta_{\mp 2},
\end{equation}
where
\begin{eqnarray}
T_{\pm 4}&=&\lambda^2(\partial_\pm\phi)^4 +(\partial_\pm^2\phi)^2 \nonumber\\
\Theta_{\pm 2}&=& -\frac{1}{2}(\partial_\pm\phi)^2\,
\frac{\partial^2V}{\partial\phi^2},
\end{eqnarray}
and \eqref{spinthree} requires
\begin{equation}
\frac{\partial^3V}{\partial\phi^3}=4\lambda^2\, \frac{\partial V}{\partial \phi}.
\end{equation}
Thus, the only possibilities are a free massive field ($\lambda=0$,
$V=m^2\phi^2/2$),
a free massless field ($V=0$, $\lambda\ne 0$), or
$$V=A e^{2\lambda \phi}+B  e^{-2\lambda \phi},$$
with $A,\, B$ being arbitrary constants.

\p
With two fields participating in different regions, the energy-like conserved
quantities will be given by the following expressions:
\begin{eqnarray}\label{energylike}
E_s&=&\int_{-\infty}^0\, dx \left(T_{s+1}^{(1)}+T_{-s-1}^{(1)}-\Theta_{s-1}^{(1)}
-\Theta_{-s+1}^{(1)}\right)\nonumber\\ &&\ \ \ \ \ \ +\int_0^\infty
dx \left(T_{s+1}^{(2)}+
T_{-s-1}^{(2)}- \Theta_{s-1}^{(2)}-\Theta_{-s+1}^{(2)}\right)\nonumber\\
&&\ \ \ \ \ \ \ \ \ \ \ \ \ \ \ \ + {\cal B}_s,
\end{eqnarray}
for a suitable boundary functional of $\phi$, ${\cal B}_s$.
The latter will be determined
by requiring
\begin{eqnarray}
\frac{dE}{dt}&=&\left(T_{s+1}^{(1)}-T_{-s-1}^{(1)}+\Theta_{s-1}^{(1)}
-\Theta_{-s+1}^{(1)}\right)_{x=0}\nonumber\\
&&\ \ \ \ \ - \left(T_{s+1}^{(2)}-
T_{-s-1}^{(2)}+ \Theta_{s-1}^{(2)}-\Theta_{-s+1}^{(2)}\right)_{x=0}\nonumber\\
&&\ \ \ \ \ \ \ \ \ \ \ \ \ \ \ \ + \frac{d{\cal B}_s}{dt} =0.
\end{eqnarray}
For the energy itself ${\cal B}_1\equiv{\cal B}$. For other values of $s$
the argument is the familiar one from \cite{Ghosh94a}, in the sense
that the existence of ${\cal B}_s$ and making use of \eqref{fieldequations}
 places severe constraints on the boundary potential ${\cal B}$.
 Thus, for $s=3$, and
after some algebra:
\begin{eqnarray}\label{conditions}
&&\lambda_1^2-\lambda_2^2=0\\
&&2\left[\lambda_1^2\left(\frac{\partial{\cal B}}{\partial\phi_1}\right)^2-
\lambda_2^2\left(\frac{\partial{\cal B}}{\partial\phi_2}\right)^2\right]
-V_1^{\prime\prime}+V_2^{\prime\prime}=0\\
&&\frac{\partial^3{\cal B}}{\partial\phi_2^2\partial\phi_1}-
\lambda_1^2\frac{\partial{\cal B}}{\partial\phi_1}
=\frac{\partial^3{\cal B}}{\partial\phi_1^2\partial\phi_2}-
\lambda_2^2\frac{\partial{\cal B}}{\partial\phi_2}=0\\
&&\frac{\partial^3{\cal B}}{\partial\phi_2^3}-
\lambda_2^2\frac{\partial{\cal B}}{\partial\phi_2}=
\frac{\partial^3{\cal B}}{\partial\phi_1^3}-
\lambda_1^2\frac{\partial{\cal B}}{\partial\phi_1}=0.
\end{eqnarray}
There are several possible solutions to these constraints. The typical one,
assuming neither of the fields is  free and massive in its bulk domain,
requires
$\lambda_1=\lambda_2=\lambda\ne 0$. Consequently
(ignoring an overall additive constant),
\begin{equation}\label{boundaryterm}
{\cal B}=a e^{\lambda(\phi_1+\phi_2)}+b e^{\lambda(\phi_1-\phi_2)}
+ce^{-\lambda(\phi_1-\phi_2)} +d e^{-\lambda(\phi_1+\phi_2)}
,
\end{equation}
where $a,b,c,d$ are constants, and the bulk potentials are given by
\begin{eqnarray}\label{bulkpotentials}
&&V_1=A_1 e^{2\lambda \phi_1}+B_1  e^{-2\lambda \phi_1}\nn\\
&&V_2=A_2 e^{2\lambda \phi_2}+B_2  e^{-2\lambda \phi_2},
\end{eqnarray}
with $A_1=2\lambda^2ab,\ A_2=2\lambda^2 ac,\ B_1=2\lambda^2 cd,\
B_2=2\lambda^2 bd$. Notice that this case allows one of the bulk fields to be
massless and free but the other need not necessarily be
(for example, taking $c=d=0$ leads to a free massless field
in $x>0$, with a Liouville field in $x<0$).

\p
The alternative is that both fields are free and massive,
so that $\lambda_1=\lambda_2=0$. In that case, the conditions on the boundary
and bulk potentials require the two masses to be the same, with a boundary
potential of the general quadratic form:
$${\cal B}=a\phi_1^2 +b\phi_1\phi_2+c\phi_2^2,$$
where $a,b,c$ are constants.

\p
If one of the bulk fields is free and massless (say $\phi_1$),
the other field
may either also be free and massless, in which case the boundary
potential has the
form
$$
{\cal B}= ae^{(\phi_1\pm\phi_2)}+de^{-(\phi_1\pm\phi_2)},
$$ or the other field may be  Liouville, in which case the boundary term
has the form
\begin{equation}\label{Liouville}
{\cal B}=\frac{m}{\beta^2}e^{\beta\phi_2/\sqrt{2}}\left(
\sigma e^{\beta\phi_1/\sqrt{2}}+\frac{1}{\sigma}
e^{-\beta\phi_1/\sqrt{2}}\right).
\end{equation}

\p
Finally, both fields could be of Liouville type. For example,
choosing $d=0$ in \eqref{boundaryterm} leads to $B_1=B_2=0,\ A_1\ne 0,\
A_2\ne 0$ and both potentials are Liouville potentials.

\p
Returning to the general case, the fields may be shifted by
a constant in each bulk
domain so that in \eqref{bulkpotentials} $A_1=B_1$ and $A_2=B_2$, the latter
in turn implying $c=\pm b,\ d=\pm a$.
It is convenient to choose $\lambda=\beta/\sqrt{2}$, in order to agree
with standard conventions for the sinh-Gordon model, and then to let
$a=m\sigma/\beta^2,\ b=m/\beta^2\sigma$. With those choices, the bulk and
boundary potentials are:
\begin{eqnarray}\label{standardcase}
&&V_1=\phantom{\pm} \frac{m^2}{\beta^2}\left(e^{\sqrt{2}\beta\phi_1}+
e^{-\sqrt{2}\beta\phi_1}
\right)\nn\\
&&V_2=\pm \frac{m^2}{\beta^2}\left(e^{\sqrt{2}\beta\phi_2}+e^{-\sqrt{2}\beta\phi_2}
\right)\nn\\
&&{\cal B}=\frac{m\sigma}{\beta^2}\left(e^{\beta(\phi_1+\phi_2)/\sqrt{2}}\pm
e^{-\beta(\phi_1+\phi_2)/\sqrt{2}}\right)+\frac{m}{\beta^2\sigma}
\left(e^{\beta(\phi_1-\phi_2)/\sqrt{2}}\pm
e^{-\beta(\phi_1-\phi_2)/\sqrt{2}}\right)
\end{eqnarray}
where, in all of these the `$\pm$' signs are strictly correlated.
(In the sinh-Gordon model the relative signs cannot be adjusted by a real shift
of one of the fields). There is a single free parameter $\sigma$
in the defect condition, which is perhaps puzzling since the half-line
boundary condition \eqref{boundaryterm} allows two free parameters.

\p
Notice, the model might be restricted to a half-line if $\phi_2$ say,
were to be set to a constant value. Then the
boundary condition satisfied by $\phi_1$ would be of the
general type \eqref{bc}.
However, typically, the constant value of $\phi_2$ would not satisfy the
equation of motion in $x>0$. On the other hand,
even if the equation of motion were
to be satisfied
with a constant $\phi_2$ in
$x>0$ (ie for real $\beta$, $\phi_2=0$),
the boundary condition for $\phi_2$  would generally not be satisfied at $x=0$.

\p
Notice too, in none of these cases
is there any reason why $\phi_1=\phi_2$ at $x=0$. Given the usual definition of
`topological' charge:
\begin{equation}
Q=\int_{-\infty}^0dx \, \partial_x\phi_1+\int^{\infty}_0 dx\,  \partial_x\phi_2
=\phi_2|_\infty -\phi_1|_{-\infty} + \phi_1|_0 -\phi_2|_0,
\end{equation}
it is clear the difference $\phi_1
-\phi_2$ measures the strength of the defect at $x=0$. In the sine-Gordon model,
where similar considerations would apply, this would appear to indicate that
topological charge need not be preserved, and hence that
a defect need not necessarily preserve soliton number.

\p It is worth noting that if eqs\eqref{fieldequations} were
satisfied simultaneously in the bulk then the `defect' conditions
with the choice of ${\cal B}$ given by \eqref{boundaryterm} would
become a B\"acklund transformation \cite{Back1882} relating the
two fields $\phi_1$ and $\phi_2$.  Indeed, a sufficient condition
for the B\"acklund transformation to work in the bulk would be:
$$\frac{\partial^2{\cal B}}{\partial\phi_1^2}=
\frac{\partial^2{\cal B}}{\partial\phi_2^2}, \quad
\left(\frac{\partial{\cal B}}{\partial\phi_1}\right)^2-
\left(\frac{\partial{\cal B}}{\partial\phi_2}\right)^2=2(V_1-V_2).$$
Clearly both of these are satisfied in all the cases mentioned above.
In the present setup, the `B\"acklund transformation' at $x=0$ represents the
boundary between two domains. This sheds an interesting new light on the
B\"acklund transformation itself.

\p
Generalising the idea, any number of defects may be represented similarly
at domain boundaries $x=x_1,\, x_2,\, \dots$, and at each boundary the defect
conditions ought to retain the same form, albeit with  different
free parameters $\sigma_i$ at each.
In the bulk, the B\"acklund transformation between two
solutions of the sine-Gordon equation generally
changes soliton number (typically adding or subtracting a soliton), which
appears to corroborate the suggestion above that a defect could allow
a change of topological charge. In addition, different domains may contain
fields of different character provided they are compatible with the
boundary condition.

\p
One interesting further point. The canonical momentum density (which is not
expected to be preserved because of the loss of translation invariance) is
given by:
\begin{equation}
P=\int_{-\infty}^0 dx\, \partial_t\phi_1\partial_x\phi_1 +
\int_0^{\infty} dx \, \partial_t\phi_2\partial_x\phi_2.\nonumber
\end{equation}
Although $P$ is not conserved, using the defect conditions
\eqref{fieldequations} it is not difficult to derive the following:
\begin{equation}\label{Pdot}
\frac{dP}{dt}=\left(-\partial_t\phi_2\frac{\partial{\cal B}}{\partial\phi_1}
-\partial_t\phi_1\frac{\partial{\cal B}}{\partial\phi_2} +\frac{1}{2}
\left(\frac{\partial{\cal B}}{\partial\phi_1}\right)^2-
\left(\frac{\partial{\cal B}}{\partial\phi_2}\right)^2- V_1+V_2\right)_{x=0}.
\end{equation}
The right hand side of \eqref{Pdot} is a total time derivative provided that
at $x=0$
\begin{equation}
\left(\frac{\partial{\cal B}}{\partial\phi_1}\right)^2-
\left(\frac{\partial{\cal B}}{\partial\phi_2}\right)^2- 2V_1+2V_2 =0,
\quad {\rm and} \quad \frac{\partial^2{\cal B}}{\partial\phi_1^2}=
\frac{\partial^2{\cal B}}{\partial\phi_1^2}\, .
\end{equation}
These conditions are precisely satisfied by the boundary term indicated in
\eqref{boundaryterm} (and indeed coincide with the conditions in the bulk
mentioned earlier for a working B\"acklund transformation).
In other words, there exists a functional
of $\phi_1,\, \phi_2$, call it ${\cal P_B}$, so that $P+{\cal P_B}$ is
conserved. There appears to be a `total' momentum which is preserved
containing bulk and defect contributions. Thus, the fields can exchange both
energy and momentum with the defect despite the lack of translation
invariance. Clearly, there is a generalisation of this idea to a collection
of defects situated at $x=x_1\, ,x_2,\, \dots$.

\section{The absence of reflection by defects}

Consider the consequences of a linearised version of \eqref{standardcase}
by setting
\begin{equation}
\phi_1=e^{-i\omega t}\left(e^{ikx}+R(k)e^{-ikx}\right), \quad
\phi_1=e^{-i\omega t} T(k)e^{ikx},\nn
\end{equation}
where $R$ and $T$ are reflection and transmission coefficients, respectively.
(Strictly speaking the fields are the real parts of these expressions; however,
in the linearised situation this is immaterial.) Imposing the defect conditions, it is convenient to set $k=2m\sinh\theta,\
\omega=2m\cosh\theta$, and $\sigma= e^p$,  to discover
\begin{equation}\label{linearised}
R(k)=0, \quad T(k)=-\frac{2i\cosh\theta-(\sigma -1/\sigma)}{2i\sinh\theta
 -(\sigma+1/\sigma)}=-i\frac{\sinh\left(\frac{\theta +p}{2}-\frac{i\pi}{4}\right)}
{\sinh\left(\frac{\theta +p}{2}+\frac{i\pi}{4}\right)} .
\end{equation}
This is surprising, given the remarks about B\"acklund transformations,
and should be compared with the results reported in \cite{Delf94}.
There, the emphasis was different. The equations expressing the
compatibility of
reflection and  transmission with the bulk factorisable S-matrix of a general
model was found to be highly constraining and required the bulk S-matrix
to satisfy $S^2=1$. In the context introduced here the question would be to find
all the defect transmission factors compatible with the sinh-Gordon
S-matrix, given that most probably there can be no reflection.

\p
To investigate what happens to a sine-Gordon soliton it is convenient to set
$\beta =1/\sqrt{2},\ m=1/2$ and to write the bulk equations and defect
condition as follows:
\begin{eqnarray}
&&x<0:\quad\partial^2\phi_1=-\sin\phi_1,\quad\\
&&x>0:\quad\partial^2\phi_2=-\sin\phi_2,\\
&&x=0:\quad \partial_x\phi_1-\partial_t\phi_2=
-\sigma\sin\left(\frac{\phi_1+\phi_2}{2}\right)-\frac{1}{\sigma}
\sin\left(\frac{\phi_1-\phi_2}{2}\right)\nn\\
&&\phantom{x=0:}\quad \partial_x\phi_2-\partial_t\phi_1=
\phantom{-}\sigma\sin\left(\frac{\phi_1+\phi_2}{2}\right)-\frac{1}{\sigma}
\sin\left(\frac{\phi_1-\phi_2}{2}\right)\label{condition}.
\end {eqnarray}
Then, a single soliton solution in the two regions has the form
(see for example \cite{Hiro80})
\begin{equation}\label{soliton}
e^{i\phi_a/2}=\frac{1-iE_a}{1+iE_a}, \quad E_a=C_a\,e^{\alpha_a x
+\beta_a t}, \quad \alpha_a^2-\beta_a^2=1,\quad a=1,2,
\end{equation}
where $C_a$ is real.
In order to be able to satisfy the conditions \eqref{condition} the
time dependence must match in the two domains ($\beta_1=\beta_2$)
and the constants $C_1,\ C_2$ are related by
\begin{equation}\label{delay}
C_2=\left(\frac{e^\theta +\sigma}{e^\theta -\sigma}\right)\, C_1,
\end{equation}
where as before it is convenient to let $\alpha_1=\alpha_2=\cosh\theta$
and $\beta_1=\beta_2=\sinh\theta$. Thus, the effect of the defect is to
delay or advance the soliton as it passes through. One curious feature is
that the defect can absorb or emit a soliton but only at a special value
of rapidity. This is most easily seen by examining \eqref{delay} and
noting that $C_2$ vanishes for $\sigma<0$ and $e^\theta = |\sigma|$,
and $C_2$ is infinite for $\sigma>0$ and $e^\theta = \sigma$.
In either case, the implication is that $\phi_2=0$ and a soliton
with this special rapidity, approaching the defect from the region
$x<0$, will be absorbed by it.

\section{Defect Lax pairs}

In this section the intention is to give an outline of the kind of
approach one
might adopt to set up Lax pairs in the presence of defects.

\p
To construct Lax pairs along the lines suggested in \cite{Bow95} it
is necessary to  separate slightly the boundary conditions in the two
regions $x<0$ and $x>0$, imposing the $\phi_1$ boundary condition
at $x=a$, and the $\phi_2$ boundary condition at $x=b>a$, and to
assume both fields are defined in  the 'overlap' region $a\le x\le b$. Since
the same framework applies to all the Toda, or affine Toda field
theories this section will be quite general. Thus, with the
same choices of coupling and mass scale as in the last section, the
defect Lax pairs for models based on simply-laced root data  are:
\begin{eqnarray}\label{Laxpairsa}
\hat a_0^{(1)}&=&a_0^{(1)}-\frac{1}{2}\theta(x-a)
\left(\partial_x\phi_1 -\partial_t\phi_2
+\frac{\partial{\cal B}}{\partial\phi_1}\right)H\nonumber\\
\hat a_1^{(1)}&=&\theta(a-x)a_1^{(1)}\nonumber\\
\hat a_0^{(2)}&=&a_0^{(2)}-\frac{1}{2}\theta(b-x)
\left(\partial_x\phi_2 -\partial_t\phi_1
-\frac{\partial{\cal B}}{\partial\phi_2}\right)H\nonumber\\
\hat a_1^{(2)}&=&\theta(x-b)a_1^{(2)},
\end{eqnarray}
where for  $p=1,2$,
\begin{eqnarray}\label{laxpairsb}
a_0^{(p)}&=& \frac{1}{2}\left[\partial_x\phi_p\cdot {\bf H}+
\sum_i\sqrt{n_i}e^{\alpha_i\cdot\phi_p/{2}}\left(\lambda E_{\alpha_i}-\frac{1}{\lambda}
E_{-\alpha_i}\right)\right] \nonumber\\
a_1^{(p)}&=&\frac{1}{2}\left[\partial_t\phi_p\cdot {\bf H}+
\sum_i\sqrt{n_i}e^{\alpha_i\cdot\phi_p/{2}}\left(\lambda E_{\alpha_i}+\frac{1}{\lambda}
E_{-\alpha_i}\right).\right]
\end{eqnarray}
 Here ${\bf H}$ are the generators in the Cartan subalgebra of the
semi-simple Lie algebra whose simple roots are $\alpha_i,\ i=1,\dots, r$, and
$E_{\pm\alpha_i}$ are the generators corresponding to the simple roots or
their negatives. If the theory is affine then the lowest root $\alpha_0=
-\sum_i n_i\alpha_i$ is appended to the set of simple roots. In either case,
affine
or non-affine, the two expressions \eqref{laxpairsb} are easily checked to
be a Lax pair (for more details about this, and further references, see \cite{Olive85}).

\p
To ensure the Lax pair defined by \eqref{Laxpairsa} really corresponds to a zero
curvature in the overlap of the two regions there should exist a group element
${\cal K}$ having the property
\begin{equation}\label{Kproperty}
\partial_t{\cal K}={\cal K}\hat a_0^{(2)}(t,b)-\hat a_0^{(1)}(t,a){\cal K}.
\end{equation}
Setting
\begin{equation}\label{}
  {\cal K}=e^{{-\bf \phi_2}\cdot{\bf H}/2}\,\bar{\cal K}\,
  e^{{\bf \phi_1}\cdot{\bf H}/2}\nonumber
\end{equation}
with $\partial_t{\cal\bar K}=0$
has the effect of removing the time derivatives from the defect term
in \eqref{Laxpairsa}, leading to
\begin{eqnarray}\label{Kbarequation}
&&\hspace{-25pt}\frac{\partial{\cal
B}}{\partial\phi_1}\cdot{\bf H}\, {\cal \bar K} +{\cal \bar K}\,
{\bf H}\cdot\frac{\partial{\cal B}}{\partial\phi_2}=\nonumber\\
&&\hspace{-10pt}\sum_i\sqrt{n_i}\left(-\lambda e^{\alpha_i\cdot
(\phi_1+\phi_2)/2}\left[{\cal\bar K},\, E_{\alpha_i}\right]+
\frac{1}{\lambda}\left(e^{\alpha_i\cdot
(\phi_2-\phi_1)/2}{\cal\bar K}E_{-\alpha_i}-e^{\alpha_i\cdot
(\phi_1-\phi_2)/2}\, E_{-\alpha_i}{\cal\bar K}\right)\right)\phantom{}
\end{eqnarray}
which is effectively an equation for both ${\cal B}$ and ${\cal \bar K}$,
see \cite{Bow95}. However, the structure of \eqref{Kbarequation} is
not the quite the same as that encountered previously.
Nevertheless, a perturbative solution can be sought of the form
\begin{equation}
{\cal \bar K}=1+\frac{k_1}{\lambda}+\frac{k_2}{\lambda^2}+\dots\, ,\nonumber
\end{equation}
 the $O(\lambda)$ terms are identically satisfied, and the other terms
 lead to the following set of expressions
\begin{eqnarray}\label{Kbits}
&&\hspace{-50pt}O(1): \quad\left(\frac{\partial{\cal B}}{\partial\phi_1}
+\frac{\partial{\cal B}}{\partial\phi_2}\right){\bf H}=-\sum_i\sqrt{n_i}
e^{\alpha_i\cdot(\phi_1+\phi_2)/2}\left[k_1,\,E_{\alpha_i}\right],\nonumber\\
&&\hspace{-50pt}O(1/\lambda):\quad  \frac{\partial{\cal B}}{\partial\phi_1}
\cdot{\bf H}\, k_1+k_1\, {\bf H}\cdot\frac{\partial{\cal B}}{\partial\phi_2}=\nonumber\\
&&\ \ \ \ \ \sum_i\sqrt{n_i}\left(-e^{\alpha_i\cdot(\phi_1+\phi_2)/2}\left[k_2,\,
E_{\alpha_i}\right]+E_{-\alpha_i}\left(e^{\alpha_i\cdot(\phi_2-\phi_1)/2}
-e^{\alpha_i\cdot(\phi_1-\phi_2)/2}\right)\right),\nonumber\\
&&\hspace{-50pt}\ \ \ \dots
\end{eqnarray}
The first of these can be satisfied for an arbitrary Toda model
provided $k_1=\sum_i\rho_iE_{-\alpha_i}$ and
$${\cal B}=\sum_i\sqrt{n_i}\rho_i\, e^{\alpha_i\cdot(\phi_1+\phi_2)/2}
+\tilde{\cal B}(\phi_1-\phi_2);$$
however, the $O(1/\lambda)$ equation does not appear to be compatible
with all choices of simple roots.
Indeed, most Toda models appear to be ruled out. For  the simplest, based on
the $A_1$ root system, a complete expression for ${\cal \bar K}$ is:
\begin{equation}
{\cal \bar K}=\begin{pmatrix}1&0\\ 0&1\end{pmatrix}
+\frac{\rho}{\lambda}
 \begin{pmatrix}0&1\\ 1&0\end{pmatrix}, \quad \rho_0=\rho_1=\rho,
 \end{equation}
 and
 \begin{equation}
 \tilde{\cal B}(\phi_1-\phi_2)=\frac{1}{\rho}\left(e^{\alpha(
 \phi_1-\phi_2)/2}+e^{\alpha(
 \phi_2-\phi_1)/2}\right).
 \end{equation}
Here, the conventions used are: $$\alpha_1=\alpha=\sqrt{2}=-\alpha_0,
\quad E_\alpha=\begin{pmatrix}0&1\\0&0\end{pmatrix}=E_{-\alpha}^\dagger,\
H=(1/\sqrt{2})\begin{pmatrix}1&\phantom{-}0\\0&-1\end{pmatrix}.$$
 In other words, it appears this style of
Lax pair can really only work in the presence of a defect for
the sinh-Gordon or Liouville models. This result is identical
with results obtained by examining the spin two conserved charges
for the models based on the $A_n$ root systems for $n\ge 2$, although
these will not be reported in detail here.

\p
It was pointed out many years ago that the possible
integrable boundary conditions
are very constrained for all Toda models apart from the sinh-Gordon
model in the sense that, in most cases, only a discrete set of parameters
may be introduced at a boundary \cite{Cor94a, Bow95}. It now appears
that defects are still more strongly constrained and generally cannot exist
in the Lagrangian form postulated in \eqref{Lagrangian}.

\vskip 1cm \noindent{\bf Acknowledgements} \vskip .5cm \noindent
EC thanks the organisers of the meeting for the opportunity to
present these ideas. One of us (CZ) is supported by a University of
York Studentship.
Another (EC) thanks the Laboratoire de Physique Th\'eorique de
l'\'Ecole Normale Sup\'erieure de Lyon for hospitality. The work
has been performed under the auspices of EUCLID - a European Commission
funded TMR Network - contract number HPRN-CT-2002-00325.



\begin{thebibliography}{99}

\bibitem{Fring93} A.~Fring and R.~Koberle,
{\em Factorized scattering in the presence of reflecting boundaries},
Nucl. Phys.  {\bf B421} (1994) 159; \hepth{9304141}.

\bibitem{Ghosh94a} S.\ Ghoshal and A.\ Zamolodchikov, {\em
Boundary $S$-matrix and boundary state in two dimensional
integrable field theory}, Int. Jour. Mod. Phys.\ {\bf A9} (1994),
3841; \hepth{9306002}.

\bibitem{Ghosh94b} S.\ Ghoshal, {\em Bound state boundary $S$-matrix
of the sine-Gordon Model}, Int. J. Mod. Phys.\ {\bf A9} (1994),
4801; \hepth{9310188}.

\bibitem{Cor94a} E. Corrigan, P.E. Dorey, R.H. Rietdijk and R. Sasaki,
\textit{Affine Toda field theory on a half line},
\plb{333}{1994}{83}; \hepth{9404108}.

\bibitem{Cor95} E. Corrigan, P.E. Dorey and R.H. Rietdijk,
\textit{Aspects of affine Toda field theory on a half-line}, Prog.
Theor. Phys. Suppl. {\bf 118} (1995) 143; \hepth{9407148}.


\bibitem{MacI95} A. MacIntyre, \textit{Integrable boundary conditions
for classical sine-Gordon theory}, J. Phys. {\bf A28} (1995) 1089;
\hepth{9410026}.

\bibitem{Bow95} P. Bowcock, E. Corrigan, P. E. Dorey and R. H. Rietdijk,
\textit{Classically integrable boundary conditions for affine Toda
 field  theories}, Nucl. Phys. {\bf B445} (1995) 469;   \hepth{9501098}.

 \bibitem{Zam79} A.B.~Zamolodchikov and Al.B.~Zamolodchikov,
\textit{Factorized S-Matrices in Two Dimensions as the Exact
Solutions of Certain Relativistic Quantum Field Theory
Models}, Ann. Phys. 120 (1979) 253.

\bibitem{Cor97}
E.~Corrigan,
{\em On duality and reflection factors for the sinh-Gordon model
with a  boundary},
Int.\ J.\ Mod.\ Phys.\ {\bf A13} (1998) 2709; \hepth{9707235}.


\bibitem{Chen00}  A.~Chenaghlou and E.~Corrigan, \textit{First order quantum
corrections to the classical reflection factor of the sinh-Gordon model},
Int. J. Mod. Phys. {\bf A15} (2000) 4417; \hepth{0002065}.

\bibitem{Cor00}
E.~Corrigan and A.~Taormina, {\em Reflection factors and a two-parameter
family of boundary bound states  in the sinh-Gordon model},
J.\ Phys.\ {\bf A33} (2000) 8739; \hepth{0008237}.


\bibitem{Delf94} G.~Delfino, G.~Mussardo and P.~Simonetti,
{\em Scattering theory and correlation functions in statistical
models with a line of defect},
Nucl. Phys. B 432 (1994) 518; \hepth{9409076}.


\bibitem{Castro-Alvaredo2002}
O.~A.~Castro-Alvaredo and A.~Fring
{\em From integrability to conductance, impurity systems},
Nucl. Phys. B649 [FS] (2003) 449; \hepth{0205076}.


\bibitem{Mintchev02}
M.~Mintchev, E.~Ragoucy and P.~Sorba,
{\em Scattering in the presence of a reflecting and transmitting impurity},
Phys.\ Lett.\ {\bf B547} (2002) 313; \hepth{0209052}.

\bibitem{Tarasov91} V.~O.~Tarasov,
\textit{The integrable initial-boundary value-problem on a
semiline - nonlinear Schr\"odinger and sine-Gordon equations},
Inverse Problems {\bf 7 (3)}  (1991) 435.

\bibitem{Back1882}A.~V.~B\"acklund,
{\em Zur Theorie der Fl\"aschentransformationen},
Math. Ann. {\bf 19} (1882) 387.

\bibitem{Hiro80} R.~Hirota, \textit{Direct methods in soliton theory},
in {\it `Solitons'}, eds R.K. Bullough and P.J. Caudrey (Berlin:
Springer 1980).

\bibitem{Olive85} D.~I.~Olive and N.~Turok,
{\em Local conserved densities and zero curvature conditions for Toda lattice
field theories},
Nucl. Phys. {\bf B257} (1985) 277.






\end{thebibliography}
\end{document}